\documentclass{iaus}
\usepackage{graphicx}

\title[Oxygen in Galactic Disk Stars] 
{Oxygen in Galactic Disk Stars: non-LTE abundances from the 777 nm O I triplet}

\author[Ramirez et al.]   
{Ivan Ramirez, Carlos Allende Prieto, \and David L. Lambert}

\affiliation{McDonald Observatory and Department of Astronomy, University of
Texas at Austin}

\pubyear{2005}
\volume{228}  
\pagerange{1--7}
\date{?? and in revised form ??}
\jname{From Lithium to Uranium: Elemental Tracers of Early Cosmic Evolution}
\editors{V. Hill, P. Fran\c{c}ois \& F. Primas, eds.}
\begin{document}

\maketitle

\begin{abstract}
Oxygen abundances for a large sample of dwarf and giant stars kinematically selected to be
part of the Galactic thin and thick disks have been determined from a non-LTE
analysis of the O~I triplet lines at 777 nm. The abundance analysis was
performed using the infrared flux method temperature scale, trigonometric
surface gravities, and accurate atomic data. Within this framework, the ionization balance of iron
lines could not be satisfied and so we adopted the iron abundances from Fe II lines only given that they are relatively less sensitive to
changes in the atmospheric parameters. We show the resulting [O/Fe] vs. [Fe/H]  relationship
and briefly discuss its implications. 
\keywords{stars: abundances, Galaxy: disk}
\end{abstract}


Systematic differences between oxygen abundance patterns of thin and thick disk
stars have been found by different authors using either the [O I] 630 nm line or LTE analyses of
the O~I 777 nm triplet in relatively small samples (e.g., Prochaska et al. 2000,
Bensby et al. 2004). These abundance trends provide tight constraints for models of
Galaxy formation and chemical evolution.

The O I triplet has not been extensively used for oxygen abundance
determinations due to strong non-LTE effects and saturation. These lines are, however, easy to
observe and are potentially useful if non-LTE corrections are
properly accounted for. Here we use spectra from McDonald Observatory, the Hobby-Eberly Telescope, and
the UVES-VLT database to determine Fe and O abundances in a sample of
about 450 disk stars. 


The thin/thick disk membership criterion we adopted is described in Mishenina
et~al. (2004). It essentially uses the Galactic space velocities $U,V,W$ of the sample stars
and a recent thin/thick disk parameterization. We derived effective temperatures using the color calibrations by Ramirez \&
Melendez (2005), which are based on the infrared flux
method (IRFM). Surface gravities were determined from accurate \textit{Hipparcos}
parallaxes and an estimate of the stellar masses from theoretical isochrones. 

Iron abundances were derived by comparing the equivalent widths of the observed
lines with those predicted by Kurucz no-overshooting models. We used about 150 Fe I and 15 Fe II
lines, all of which have transition probabilities measured in the laboratory and
damping constants theoretically computed as in Barklem et al. (2000). A similar approach was used to derive LTE oxygen
abundances. Restricted non-LTE corrections for the oxygen triplet were calculated using
TLUSTY (Hubeny \& Lanz 1995) and the oxygen model atom of Allende Prieto et al. (2003).

\medskip

With the ingredients described above for the Fe abundance determination,
we were not able to obtain the same mean Fe abundance from Fe I and Fe II lines. The Fe II
abundances are larger by about 0.07 dex for F and 
early G dwarfs and giants. The discrepancy worsens for late-type and metal-rich dwarfs but for giants
the difference remains constant. This inconsistency may be revealing a
systematic error in the IRFM effective temperature scale, non-LTE effects on Fe line
formation, limitations of classical model atmospheres, or a combination of them. We adopted the Fe II abundance as our [Fe/H] indicator due to
its relatively weak dependence on effective temperature (most of our sample stars are hotter
than about 5400 K) and small predicted non-LTE effects.

\begin{figure}
 \centering
 \includegraphics[bb=8 384 595 750,width=11cm]{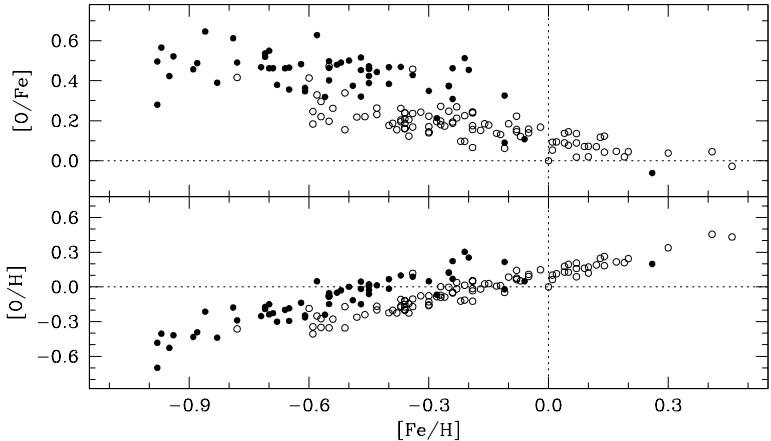}
  \caption{Oxygen abundance trends for our sample dwarf stars with 5400~K$<T_\mathrm{eff}<6000$~K and thin/thick disk membership probabilities larger than 80\%. Filled circles: thick disk stars, open
  circles: thin disk stars. The dotted lines correspond to solar values.}
  \label{f:f2}
\end{figure}

Our oxygen abundance trends are shown in Fig.~\ref{f:f2} for the dwarf stars with
5400~K$<T_\mathrm{eff}<6000$~K and thin/thick disk membership probabilities larger than
80\% (this is about 1/3 of the total sample). In the range $-0.8<$[Fe/H]$<-0.2$, thick disk stars have larger [O/Fe] and [O/H] compared to
thin disk stars. The thick disk [O/Fe] ratio is nearly constant at about 0.5
while the thin disk [O/Fe] ratio slowly declines towards solar values.

Note that a few kinematically selected thick disk stars seem to follow the thin disk
abundance pattern, and viceversa. This suggests that the kinematic criterion alone is
incomplete. In fact, if an abundance criterion is considered, no thick disk
stars would probably be found in our sample at [Fe/H]$>-0.2$.

\medskip

The fact that the oxygen abundance trends are well separated suggests that thin
and thick stars were formed from different mixtures of gas. Along with other Galactic and extragalactic
evidence, the merger scenario for thick disk formation has been favored (see Bensby et al. 2004 for details and
references). An interesting result is that we do not find many high metallicity thick disk 
stars and we do not obtain a decline in the [O/Fe] ratio of thick disk stars,
attributed to SNIa contribution to the gas that formed thick disk stars by
Bensby et~al. (2004). Since our sample selection did not include a [Fe/H] criterion,
this may be indicating an upper limit to the [Fe/H] distribution of thick disk stars at
[Fe/H]$\sim-0.2$. Ongoing observations will help us to clarify this important matter.


\end{document}